\documentclass[sn-mathphys]{sn-jnl}
\usepackage[utf8]{inputenc}
\usepackage{times}
\usepackage{t1enc}
\usepackage{pdftricks}
\usepackage{mathrsfs}
\usepackage{mathtools}
\usepackage{amsmath,amssymb,float,amsthm}

\newcommand{\tr}{\mathrm{Tr}}

\newcommand{\cB}{\mathcal{B}}
\newcommand{\cC}{\mathcal{C}}

\newcommand{\cE}{\mathcal{E}}
\newcommand{\cF}{\mathcal{F}}

\newcommand{\cH}{\mathcal{H}}
\newcommand{\cI}{\mathcal{I}}

\newcommand{\cP}{\mathcal{P}}

\newcommand{\cR}{\mathcal{R}}

\newcommand{\cW}{\mathcal{W}}

\DeclareMathOperator*{\argmax}{argmax}

\theoremstyle{thmstyleone}%
\theoremstyle{thmstyletwo}%

\theoremstyle{thmstylethree}%

\begin{document}

\title[QEC: Noise-adapted Techniques and Applications]{Quantum Error Correction: Noise-adapted Techniques and Applications}

%\title{Current trends in quantum error correction}

\author*[1]{\fnm{Akshaya} \sur{Jayashankar}}\email{aksh29@gmail.com}

\author[2]{\fnm{Prabha} \sur{Mandayam}}\email{prabhamd@physics.iitm.ac.in}

\affil*[1]{\orgdiv{Department of Physics}, \orgname{Indian Institute of Technology Madras}, \orgaddress{\city{Chennai}, \postcode{600036}, \country{India}}}

\affil[2]{\orgdiv{Department of Physics}, \orgname{Indian Institute of Technology Madras}, \orgaddress{\city{Chennai}, \postcode{600036}, \country{India}}}

\date{\today}

%\begin{document}

\abstract{
The quantum computing devices of today have tens to hundreds of qubits that are highly susceptible to noise due to unwanted interactions with their environment. The theory of quantum error correction provides a scheme by which the effects of such noise on quantum states can be mitigated, paving the way for realising robust, scalable quantum computers. In this article we survey the current landscape of quantum error correcting (QEC) codes, focusing on recent theoretical advances in the domain of noise-adapted QEC, and highlighting some key open questions. We also discuss the interesting connections that have emerged between such adaptive QEC techniques and fundamental physics, especially in the areas of many-body physics and cosmology. We conclude with a brief review of the theory of quantum fault tolerance which gives a quantitative estimate of the physical noise threshold below which error-resilient quantum computation is possible.}


\maketitle

\section{Introduction}

Quantum computing technologies have advanced by leaps and bounds over the last decade. We are already witness to the first generation of quantum processors successfully demonstrating that the theoretical promise of quantum computational speedups can be realised in practice~\cite{qSupremacy, boson_sampling2021}. There are however a few key challenges that must be overcome in order to scale from the current generation of noisy intermediate-scale quantum (NISQ) devices~\cite{preskill2018_quantum}, to robust universal quantum computers. Error mitigation and fault tolerance is arguably the biggest of these challenges, both from a theoretical and experimental standpoint.

The theory of quantum error correction (QEC)~\cite{lidar} lays down the basic framework for dealing with noise affecting the  quantum states of interest. The early works of Shor~\cite{Shor95, Calderbank_shor96} and Steane~\cite{Steane97} demonstrated how quantum error correcting codes can be constructed for bit-flip and phase-flip noise, by making use of entanglement to circumvent the challenge posed by no-cloning theorem~\cite{no_cloning82}. Subsequently, a general theory of QEC was developed for arbitrary errors, in terms of algebraic~\cite{knill97} and information-theoretic conditions~\cite{infoqec_96}. Today, the standard approach to quantum error correction works by discretizing the errors affecting the quantum state in terms of  the \emph{Pauli basis}~\cite{nielsen}, with the group structure of the Pauli operators naturally leading to the rich mathematical framework of quantum stabilizer codes~\cite{gottesman1997}. We refer to a recent review~\cite{terhal2015} that surveys the major developments in QEC, from the stabilizer codes and CSS codes to the more recently proposed classes of topological codes~\cite{bombin2013} and surface codes~\cite{raussendorf2007}.

In contrast to the \emph{general-purpose} QEC schemes described above, error correction protocols tailored to specific noise models have been devised, starting with a $4$-qubit code that protects against amplitude-damping noise~\cite{leung}. Such adaptive QEC protocols have shown to offer the same degree of protection while using fewer resources~\cite{hui_prabha, fletcher_AD}, particularly in the case of non-Pauli noise models. In this article, we focus on this emerging area of \emph{channel-adapted} or \emph{noise-adapted} QEC and survey some of the recent progress in this area. We will first review the early theoretical progress on approximate QEC~\cite{fletcher_rec, beny2010, tyson2010, mandayam2012}, and introduce an important noise-adapted recovery map, namely, the Petz map~\cite{barnum, Petz}. We then discuss various recent approaches to constructing noise-adapted quantum codes~\cite{ak_cartan, qvector, cao2022quantum, reinforcement}. Next we survey some interesting applications of noise-adapted recovery maps, especially in the context of many-body quantum systems. We also briefly touch upon the interesting role that approximate recovery maps are coming to play in the AdS/CFT setting~\cite{holography_qec}. 

While QEC focuses on dealing with the noise affecting the quantum states assuming that all the gate operations are ideal, the theory of quantum fault tolerance~\cite{preskill1998} lays down the basic framework for dealing with faulty gate operations in quantum computing devices. Fault-tolerant protocols build upon the QEC framework and enable the construction of quantum circuits that are truly noise-resilient, provided the noise is below a certain \emph{threshold} value~\cite{knill2005, aliferis2006}. We devote the final section of our review to discussing the latest developments in the theory of quantum fault tolerance, focusing on fault-tolerant schemes that are based on noise-adapted codes~\cite{ak_ft}.

The rest of this review is organized as follows. In Sec.~\ref{sec:perfect_aqec}, we introduce the basic mathematical formalism of perfect and approximate QEC. In Sec.~\ref{sec:noise-adapted} we formally introduce the idea of noise-adapted codes and recovery maps. We then proceed to discuss various analytical and numerical approaches to constructing such noise-adapted recovery maps and noise-adapted quantum codes. In Sec.~\ref{sec:applications}, we survey some interesting applications of adaptive QEC protocols. In Sec.~\ref{sec: FT} we discuss how quantum fault tolerance can be achieved using noise-adapted QEC protocols. We conclude with a brief summary and future outlook in Sec.~\ref{sec:summary}.

\section{Preliminaries}\label{sec:perfect_aqec}

%: Perfect and Approximate Quantum Error Correction

\begin{figure}[H]
\centering
\includegraphics[scale=.5]{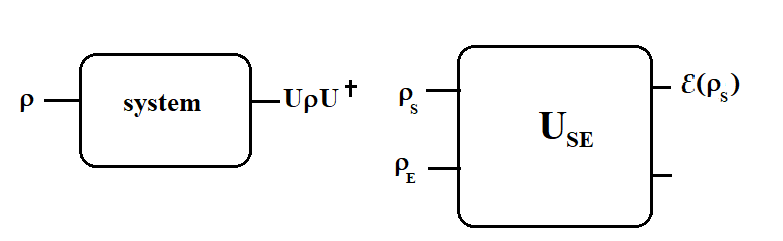}
\caption{ Unitary evolution of a closed system (left) vs non-unitary dynamics of an open quantum system (right)~\cite{thesis}.}
\label{fig:quantum_channel}
\end{figure}

We begin with a brief review of the mathematical description of decoherence in quantum systems, and refer to some of the well-known textbooks~\cite{nielsen, lidar} for further details. Noise in quantum systems is described by the mathematical framework of quantum operations, often referred to as \emph{quantum channels} in the context of quantum computation~\cite{nielsen}. A quantum channel is a completely positive trace-preserving (CPTP) map describing the non-unitary evolution of an open quantum system interacting with its environment, as shown in Fig.~\ref{fig:quantum_channel}. The action of such a map $\cE$ on the state $\rho$ of a quantum system can be described via a set of \emph{Kraus operators} $\{E_{i}\}$, as, 
$\cE(\rho) = \sum_{i}E_{i}\rho E_{i}^{\dagger}$, satisfying the normalization condition,  $sum_i E_i^\dagger E_i = I$. These Kraus operators essentially represent the different \emph{error operators} associated with the noise channel $\cE$. 

\begin{figure}[H]
\centering
\includegraphics[scale=.7]{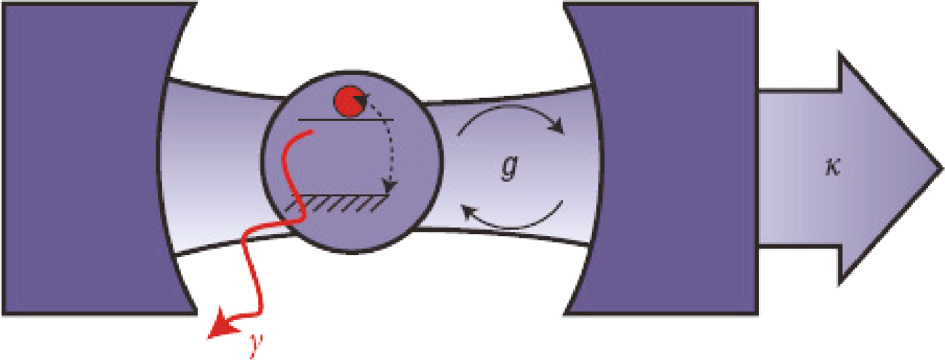}
\caption{ An atom in a cavity undergoing spontaneous emission~\cite{cao}.}
\label{fig:atom_cavity}
\end{figure}

A classical example of  quantum channel arising out of a system-environment interaction is the \emph{amplitude-damping channel}. This is a channel that models dissipation in a two-dimensional quantum system (a qubit!), such as a two-level atom interacting with an optical mode in a cavity via the Jaynes-Cummings Hamiltonian, as shown in Fig.~\ref{fig:atom_cavity}. The action of the amplitude-dampng channel on the qubit system is described by a a pair of Kraus operators, $E_{0}$ and $E_{1}$ which are given by~\cite{nielsen},

\begin{equation}\label{eq:ampdamp}
E_{0}  = \vert0\rangle\langle 0\vert+ \sqrt{1-\gamma}\,\vert1\rangle\langle 1\vert, \; \quad E_1 = \sqrt{\gamma}\,\vert0\rangle\langle 1\vert.
\end{equation}
Here, $\gamma$ is the probability of decay of the excited state $\vert1\rangle$ to the ground state $\vert0\rangle$. Amplitude-damping noise is an important model of \emph{non-Pauli} noise, since its Kraus operators are not simply proportional to Pauli operators. This is in contrast to other well known noise models such as bit-flip,  phase-flip and depolarizing noise which are all described by Kraus operators that belong to the set  $\{I,X,Y,Z\}$ of single-qubit Pauli operators.

\subsection{Quantum Error Correction} \label{sec:qec}

Quantum error correction involves protecting the information in say, a $d_0$-dimensional Hilbert space $\cH_0$, by \emph{encoding} it in a $d_0$-dimensional subspace $\cC$ of a larger Hilbert space $\cH$. Subsequently, a \emph{recovery} map $\cR$ is applied in order to reverse the effects of errors, followed by \emph{decoding}, which brings the information back into the original Hilbert space $\cH_0$. The encoding $\cW$ is an isometry whose action can be described as, $\cW:\cB(\cH_0)\rightarrow \cB(\cC)\subseteq\cB(\cH)$, where $\cB(.)$ denotes the set of bounded linear operators on the respective Hilbert spaces. The subspace $\cC$ is called the \emph{codespace} or simply, a quantum code~\cite{nielsen, lidar}. 

\begin{figure}[H]
\centering
\includegraphics[scale=.25]{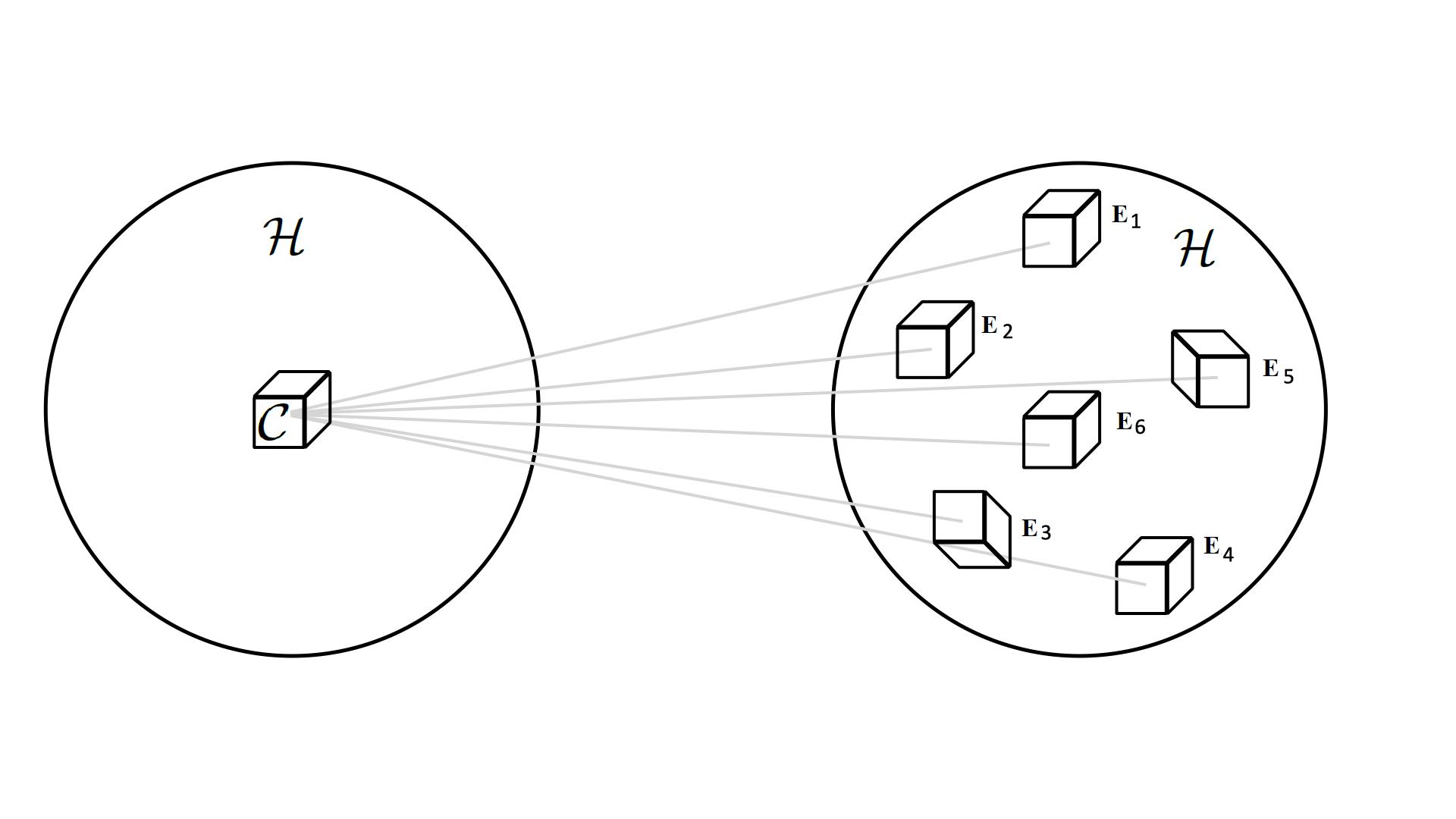}
\caption{Perfect QEC: codespace gets mapped to mutually orthogonal subspaces by the errors $\{E_i\}$~\cite{thesis}. }
\label{fig:perfectqec}
\end{figure}

Ideally, to protect against the error $\{E_{i}\}$ caused by a quantum channel $\cE$, the codespace must be chosen such that it %become distinguishable under the action of the Pauli errors. They 
gets mapped onto mutually orthogonal subspaces under the action of the errors, as shown in Fig.~\ref{fig:perfectqec}. This allows for the errors $\{E_{i}\}$ afflicting the codespace to be identified unambiguously and hence corrected for. Such codes are referred to as \emph{perfectly} correctable codes. This property of a quantum code to be perfectly correctable under the action of the errors $\{E_{i}\}$ is captured by the Knill-Laflamme conditions, stated as~\cite{knill97},
\begin{equation}\label{eq:perfectqec}
 PE_{i}^{\dagger}E_{j}P = \alpha_{ij}P, \; \forall i, j. 
\end{equation}

\noindent Here, $P$ is the projection onto the codespace $\cC$ and $\alpha_{ij}$ are complex elements of a Hermitian matrix $\alpha$. The algebraic conditions in Eq.~\eqref{eq:perfectqec} are necessary and sufficient for the existence of a recovery (CPTP) map $\cR$ which can correct perfectly for the errors $\{E_i\}$. We note here that information-theoretic conditions for perfect QEC have also been formulated in terms of the coherent information~\cite{infoqec_96}. In contrast to the active approach to QEC involving a recovery operation, there exist passive QEC strategies~\cite{lidar}, where the information to be protected is stored in subspaces or subsytems of a larger Hilbert space that remain unaffected by the noise, {called Decoherence-free subspaces (DFS) and Noiseless subsystems (NS)~\cite{kribs}, respectively}. Such passive QEC schemes are outside the scope of this review, here we will focus exclusively on active quantum error correction. 

Since the QEC conditions in Eq.~\eqref{eq:perfectqec} are linear, a code that corrects perfectly for the set $\{E_{i}\}$ will also correct perfectly for any linear combination of the errors in the set. Thus, to correct for an arbitrary single-qubit error, it suffices to find codes that can correct perfectly for a unitary error basis such as the Pauli basis spanned by the single-qubit Pauli operators $\{I,X,Y,Z\}$. This in turn leads to an elegant reformulation of the theory of perfect QEC using the group theoretic algebra of Pauli operators called the stabilizer formalism~\cite{lidar,nielsen}. We refer to other review articles in this special issue for further details on the stabilizer formalism and its role in quantum code construction.

Based on the set of errors correctable by it, a quantum code is characterized by three parameters. An $[[n, k, d]]$ quantum code is one the encodes $k$ qubits into $n$ physical qubits and corrects for errors on upto $t$ qubits, where, $d=2t+1$ is often referred to as the distance of the code. Stabilizer codes such as the well known $9$-qubit Shor code~\cite{Shor95}, the $7$-qubit Steane code~\cite{Steane97} and the $5$-qubit code~\cite{laflamme_96} are all general-purpose quantum codes, since they can correct for arbitrary single-qubit errors. However, their error-detection and error-correction mechanisms work by discretizing arbitrary errors in terms of the Pauli errors and distinguishing single-qubit Pauli errors perfectly. These perfect QEC codes are in  essence tailored to protect against single-qubit Pauli noise, which in turn imposes rigid constraints on the structure of such codes. In what follows, we will look at how it is possible to construct newer classes of quantum codes both by deviating from the demands of perfect QEC and by moving beyond the class of Pauli noise channels.

%Recall that a set of $n$-fold tensor product Pauli operators form a group $\mathscr{G}_{n}$ as given below.
%\[\mathscr{G}_{n}= \{\pm I^{\otimes n}, \pm i I^{\otimes n}, \pm X I^{\otimes n-1}, \pm i X I^{\otimes n-1}, \pm Y I^{\otimes n-1}\ldots, \pm i Z^{\otimes n}\} .\]
%where $\{I,X,Y,Z\}$ are single-qubit Pauli operators. A stabilizer group $\mathscr{S}$ is defined as an abelian subgroup of $\mathscr{G}_{n}$, with tensor product Pauli operators as its elements that commute with each other. This set necessarily excludes the element $-I$. The common eigenspace shared by the set of commuting elements of the stabilizer $\mathscr{S}$ define the codespace.
%Any Pauli operator in the group $\mathscr{G}_{n}$ either commutes or anticommutes with a stabilizer group element $\mathscr{S}$. An error, say $B$, which is a Pauli operator element outside the group $\mathscr{S}$ necessarily anticommutes with atleast one element, say $A$ in $\mathscr{S}$, such that $B A =-A B$. Then the action of the error $B$ on the codespace $\cC$ is given by,
%\begin{equation}\label{eq:stabilizer}
%A(B \vert v\rangle) =- B( A \vert v\rangle) = -B\vert v\rangle , \forall \vert v\rangle \in \cC.
%\end{equation}
% which would mean that the error, $B$ affecting the codespace  can be detected distinctly when one measures $A$, since $B$ has now unitarily rotated the codespace  into a subspace which is $-1$ eigenvalue eigenspace of operator $A

\subsection{Approximate Quantum Error Correction}\label{sec:aqec}

In order to correct perfectly for arbitrary single-qubit errors by decomposing them in terms of Pauli errors, a single qubit must be encoded into at least five qubits~\cite{knill97}. In violation of this quantum Hamming bound, the pioneering work of Leung \emph{et al.}~\cite{leung} provided an example of $4$-qubit QEC code that can correct for single-qubit amplitude-damping noise. %and is not distinguishable in the standard Pauli $\{I, X, Y, Z\}$ error basis. 
It was further shown that this $4$-qubit code does not satisfy the Knill-Laflamme conditions in Eq.~\ref{eq:perfectqec} exactly, but satisfies a \emph{perturbed} form of the QEC conditions. The $4$-qubit code is thus the first example of an \emph{approximate} quantum code, with a codespace that is spanned by,
\begin{equation}\label{eq:4qubit}
\vert0_L\rangle = \tfrac{1}{\sqrt 2}(\vert0000\rangle +\vert1111\rangle), \quad
\vert1_{L}\rangle = \tfrac{1}{\sqrt 2}(\vert1100\rangle +\vert0011\rangle).
 \end{equation}
Finally, it was shown that this approximate $4$-qubit code achieves a comparable degree of protection as the standard $5$-qubit code, against amplitude-damping noise~\cite{leung}. These observations generated a lot of interest in looking for approximate codes and QEC protocols.

\begin{figure}[H]
\centering
\includegraphics[scale=.25]{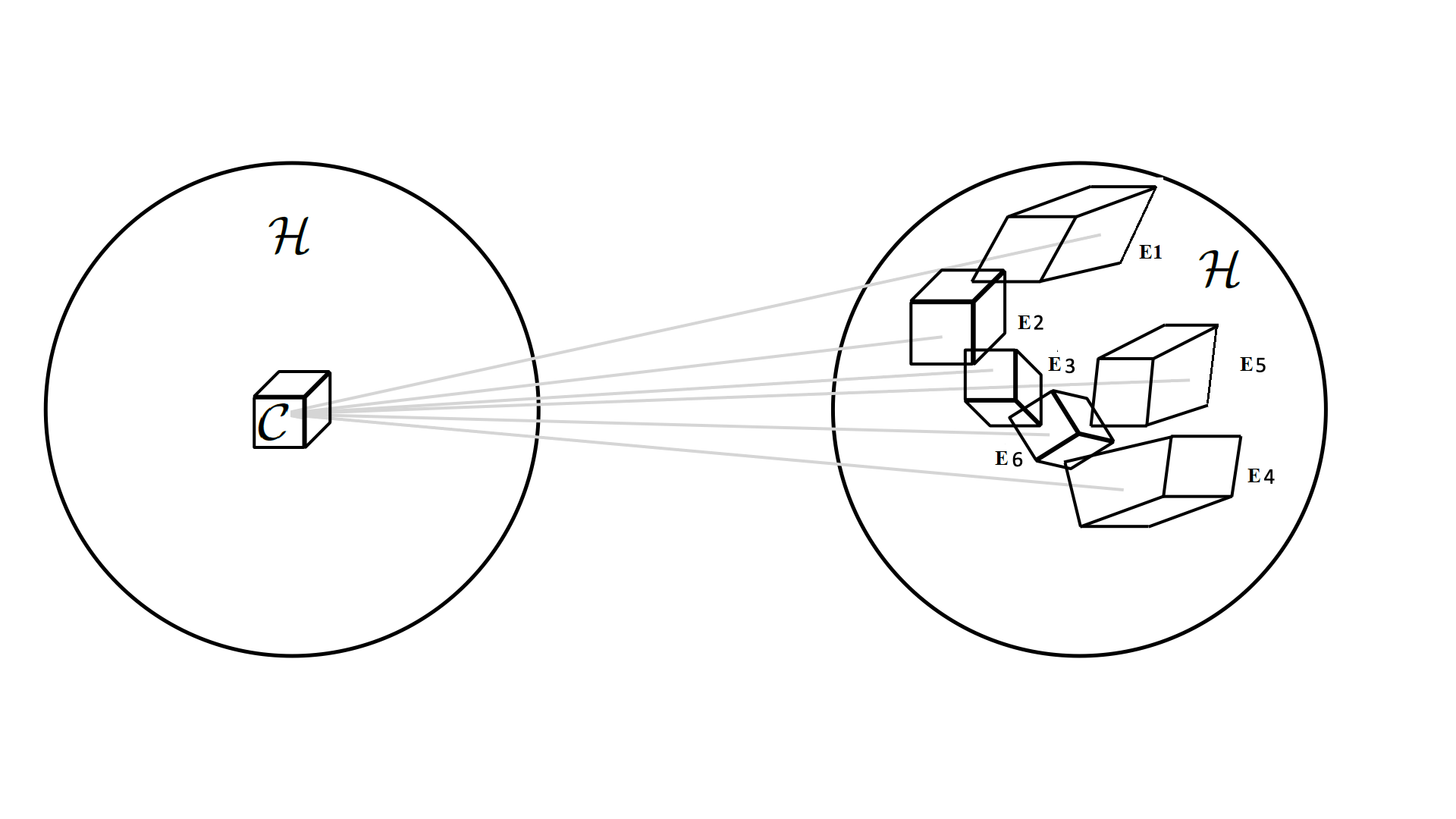}
\caption{Approximate QEC: codespace gets mapped to overlapping subspaces due to errors $\{E_i\}$~\cite{thesis}. }
\label{fig:approxqec}
\end{figure}

In contrast to the idea of perfect QEC, approximate QEC is achieved by relaxing the demands on the code. The codespace need not be mapped to distinct subspaces under the action of the errors, but can instead be mapped to overlapping subspaces as shown in Fig.~\ref{fig:approxqec}. Furthermore, while the standard recovery procedure for a perfect code simply involves applying the appropriate Pauli operator once the error is detected and identified, an approximate code often requires a non-trivial recovery, one that is also adapted to the noise, so as to reverse the effects of the noise in the best possible way. Thus, we may also refer to approximate codes and approximate QEC protocols as \emph{noise-adapted} schemes, since the code and/or the recovery are tailored to a particular noise model. 

Formally, a channel $\cE$ is approximately correctible on codespace $\cC$ if and only if there exists a CPTP map $\cR$ such that $\mathbb{F}[\cC; \cR \circ \cE] \approx 1 $, where $\mathbb{F}$ is any average or worst-case fidelity measure, quantifying how well the combination of $\cE$ followed by $\cR$ protects the states in the codespace. Before we proceed with our survey of approximate or noise-adapted QEC, it may be useful to recall two standard fidelity measures that are used to quantify the performance of any QEC protocol. The first is the \emph{entanglement fidelity} $F_{e}(\rho, \cE)$, which describes how well the noise channel $\cE$ preserves entanglement. Specifically, let $\vert\Psi_{AB}\rangle$ be a purification of the state $\rho_{A}$ onto the the joint system $\cH_{A}\otimes\cH_{B}$. Then, $F_{e}(\rho, \cE)$ defined as~\cite{nielsen},
 \begin{equation}\label{eq:entfid}
     F_e(\rho, \cE) = \langle \Psi_{AB}\vert (\cE_{A}\otimes \cI_{B}) (\vert \Psi_{AB}\rangle\langle\Psi_{AB} \vert)\vert \Psi_{AB}\rangle,
 \end{equation}
where the noise $\cE_{A}$ acts only on the subsystem $\cH_A$. In contrast to this, the \emph{fidelity} $F (\rho,\cE(\rho))$ quantifies how well the channel preserves the state $\rho$ in terms of the Bures' metric, and is defined as~\cite{nielsen},
\begin{equation}
    F (\rho, \cE) = \tr\sqrt{\sqrt{\rho}\cE(\rho)\sqrt{\rho}} \label{eq:fid}
\end{equation}
The performance of a QEC protocol described by the pair of encoding and recovery maps $(\cW,\cR)$ for a noise $\cE$ is then quantified by the \emph{worst-case fidelity}, defined as,
\begin{equation}
F_{\min}(\cW, \cR;\cE) \equiv \min _{\vert \psi\rangle \in \cH_{0}} F(\vert \psi\rangle,\cW^{-1}\circ \cR \circ \cE \circ \cW). \label{eq:worstcase_fidelity}
\end{equation}

%\textcolor{blue}{Works of Barnum-Knill, Beny-Oreshkov, Hui-Prabha}

%\textcolor{blue}{Figures of merit for AQEC: Entanglement Fidelity and worst-case fidelity}

\section{Noise-adapted Quantum Error Correction}\label{sec:noise-adapted}

Much of the existing work on QEC focuses on general-purpose QEC codes that assume no knowledge about the noise affecting the system. This allows one to construct codes capable of correcting any arbitrary error by discretizing the error in terms of Pauli operators, as explained earlier. However, such Pauli-based QEC codes are resource intensive, whereas noise-adapted codes like the $4$-qubit code demonstrated that it may be possible to achieve a comparable degree of protection using fewer qubits. The problem of noise-adapted QEC is to identify a pair $(\cW,\cR)$ of encoding and recovery maps respectively, that achieve optimal protection against the noise map under consideration.  

For a given noise map $\cE$ (assuming some dimension $d$ of the physical system), the best QEC protocol is thus the solution to an optimization over encoding isometries $\cW$ and recovery maps $\cR$, for a chosen measure of fidelity $\mathbb{F}$.
\begin{equation}
    \argmax_{\cW}\argmax_{\cR} \mathbb{F}(\cW, \cR; \cE). \label{eq:opt}
\end{equation}
This is in general a hard problem, since it involves a double optimization or a triple optimization depending on whether the measure $\mathbb{F}$ is an average fidelity measure or a worst-case fidelity measure such as the one defined in Eq.~\eqref{eq:worstcase_fidelity}. The problem of noise-adapted QEC becomes more tractable when these two tasks are decoupled: we may instead ask what is the best possible encoding assuming a fixed recovery map, or, we may try to solve for the best recovery map for a given encoding. In what follows, we will first survey the known results on noise-adapted recovery maps (Sec.~\ref{sec:adaptive_recovery}) and then proceed to discuss different approaches to search for and construct good noise-adapted quantum codes (Sec.~\ref{sec:adaptive_codes}).

%construct  and require a lot of physical qubits to protect even a single qubit worth information %, for instance, the non-degenerate $[[5,1,3]]$ code, Steane's $ [[7,1,3]] $code. In the current day and age where noisy-intermediate-scale quantum (NISQ) devices showcase a few hundred noisy qubits and noisy gates, performing scalable computation becomes a serious challenge. One approach to sustaining long and reliable computations  is then to characterise the noise affecting the system and then perform QEC specific to this noise, which is called \emph{adaptive} QEC. Majority of the work along this thread rests on the fact that the noise model and the QEC code is Pauli-based, for example, QEC for dephasing noise based on $3$-qubit code. However,
 
 \subsection{Noise-adapted recovery maps}\label{sec:adaptive_recovery}
 
One of the earliest analytical results on noise-adapted QEC was the demonstration of the existence of a universal, near-optimal recovery map for any noise channel, with optimality defined in terms of the average entanglement fidelity~\cite{barnum}. This noise-adapted recovery map is based on an original construction due to Petz~\cite{petz2003}, and is often referred to as the \emph{Petz map} today in the literature. Subsequently, this idea was used to obtain conditions for approximate QEC~\cite{beny} as a generalization of the Knill-Laflamme conditions, which in turn provided a way to construct a near-optimal recovery map in terms of the worst-case entanglement fidelity. In related work, it was shown that a modified version of the Petz map is in fact a noise-adapted recovery map for any combination of noise and codespace, achieving near-optimal worst-case fidelity~\cite{hui_prabha}, leading to simple algebraic conditions for approximate QEC as a perturbation of the Knill-Laflamme conditions. We note here that information-theoretic conditions for approximate QEC have also been formulated~\cite{SW2002}, as a perturbation of the perfect conditions.

Given the important role played by the Petz map, it might be useful to briefly review its definition and properties here. The map was originally introduced in an information-theoretic setting, in the context of saturating the monotonicity of the quantum relative entropy~\cite{Petz}. For a density operator $\rho$ and noise channel $\cE$ with Kraus operators $\{E_{i}, i = 1,2, \ldots, N\}$, the Petz map is defined via its Kraus operator decomposition as follows.
\begin{equation}\label{eq:petz1}
    \cR_{\rho} \sim \lbrace R_i \equiv \sqrt{\rho} E_{i}^{\dagger} \cE(\rho)^{-1/2} \rbrace .
\end{equation}
Note that this map recovers the state $\rho$ perfectly after the action of the noise $\cE$. This \emph{state-specific} Petz map was generalized for an ensemble of states in~\cite{barnum}, to obtain a near-optimal recovery map in terms of the average entanglement fidelity.

Subsequently, a \emph{code-specific} Petz map $\cR_{P}$ was defined for the noise channel $\cE$ with Kraus operators $\{E_{i}, i = 1,2, \ldots, N\}$ and a codespace $\cC$, as,
\begin{equation}
    \cR_{P} \equiv \{ PE_{i}^{\dagger}\cE(P)^{-1/2} \}, \; i = 1,2,\ldots, N, \label{eq:petz_kraus}
\end{equation} 
where, $P$ is the projector onto the codespace and $\cE(P)=\sum_{i}E_{i}PE_{i}^{\dagger}$. The action of the code-specific Petz map is thus defined on the support of $\cE(P)$, as,
\begin{equation}\label{eq:Petzmap}
\cR_{P}(\cdot) \equiv \sum_{i=1}^{N} PE_{i}^{\dagger} \cE(P)^{-1/2}(\cdot) \cE(P)^{-1/2}E_{i}P.
\end{equation}
Note that $\mathcal{R}_{P}$ can be thought of as being composed of three CP maps as follows.
\begin{align}
  \mbox{Normalizer map :} &\quad \cE(\cP_{\cC})^{-1/2}(.)\cE(\cP_{\cC})^{-1/2} \nonumber \\
  \mbox{Adjoint map :}  &\quad \cE^{\dagger} (.) \nonumber \\
  \mbox{Projector map :}  &\quad \cP_{\cC}\,(.) \cP_{\cC} \label{eq:petz3}
\end{align}
This map was shown to achieve close to optimal worst-case fidelity for the codespace $\cC$ affected by the noise channel $\cE$~\cite{hui_prabha}. Furthermore, in case the codespace $\cC$ is perfectly correctable for noise $\cE$, it can be shown that the map $\cR_{P}$ is indeed the standard recovery map of perfect QEC! We note here that the code-specific Petz map construction was further generalised as a near-optimal recovery for subsystem codes~\cite{mandayam2012}. On a related note, approximate QEC has also been illustrated in a different setting, where the noise channel is assumed to act jointly on system and bath~\cite{hui_2}.

In recent times, the Petz map which is a close analogue of the classical Bayesian reversal map~\cite{petz_bayesian} has been used to recover information in the case of non-Markovian dynamics~\cite{petz_nonmarkovian}. A physical protocol which implements the Petz map in the form of Hamiltonians and jump operators has been given in Ref.~\cite{petz_physicalprocess} . The Petz map construction for the case of continuous variable case, namely the bosonic Gaussian channel has also been studied in Ref.~\cite{wilde}. Finally, a quantum algorithm to implement Petz map based on the quantum singular value decomposition was proposed recently~\cite{gilyen2022}.

We conclude this section with a summary of numerical studies of optimal noise-adapted recovery maps. Solving for the optimal recovery map in terms of the worst-case fidelity is computationally hard given that Eq.~\eqref{eq:opt} requires a triple optimization in this case. However, upon relaxing certain constraints, it was shown that this triple optimization is tractable via a semidefinite program (SDP), although the recovery map thus obtained is typically suboptimal~\cite{yamamoto}. The entanglement fidelity measure defined in Eq.~\eqref{eq:entfid}, on the other hand, is amenable to convex optimization techniques~\cite{kosut}. In terms of the worst-case entanglement fidelity, the problem of finding the optimal noise-adapted code for a fixed recovery and the problem of finding the optimal noise-adapted recovery for a fixed code can both be recast as semidefinite programs, which are simply dual to each other~\cite{fletcher_rec, fletcher_thesis}. 
%structured recovery adapted to the noise process considered Optimal noise-adapted codes and noise-adapted recovery maps have been identified numerically via other convex optimization techniques, fletcher_rec}, using the entanglement fidelity as the figure of merit.  

\subsection{Noise-adapted quantum codes}\label{sec:adaptive_codes}

Starting with the $4$-qubit code~\cite{leung} described in Eq.~\eqref{eq:4qubit} above, several analytical constructions of approximate codes have been obtained in the literature. These include the class of \emph{cat-codes}~\cite{catcode1, catcode2} and \emph{binomial codes}~\cite{binomial2016} which are tailored towards protecting information stored in bosonic modes against dominant noise processes such as photon-loss and amplitude-damping. There are a few analytical constructions of noise-adapted codes for amplitude-damping noise~\cite{shor_lang2007, shor2011} and generalized amplitude-damping~\cite{cafaro2014}, which make use of the structure of the Kraus operators of the respective channels to get the desired error mitigation properties.

Constructing noise-adapted codes for an arbitrary quantum channels is in general a hard problem. Recently, a systematic way of constructing optimal noise-adapted approximate codes for general noise models using the Cartan decomposition was provided in Ref.~\cite{ak_cartan}. Furthermore, in contrast to earlier numerical approaches, the worst-case fidelity was used as the figure of merit to obtain the optimal codes. We briefly summarize this approach here before surveying other numerical approaches to finding noise-adapted quantum codes.

As a first step towards simplifying the search for good quantum codes, the recovery map is simply fixed to be the code-specific recovery map $\cR_P$ defined in Eq.~\ref{eq:Petzmap}. Specifically, for a qubit noise channel $\cE$ and two-dimensional code $\cC$, using the Petz map leads to a simple optimization problem for the worst-case fidelity, namely~\cite{ak_cartan},
\begin{equation}\label{eq:fid_loss}
 F_{\rm min} (\cW) = 1- \frac{1}{2}{\left[1-t_{\min}(\cW)\right]}.
\end{equation}
where $t_{\rm min}$ refers to the smallest eigenvalue of a $3\times 3$ matrix. Therefore, the encoding unitary $\cW$ which fixes the optimal codespace is the one that maximizes the fidelity, or, minimizes the fidelity loss $\eta_\cW = 1 -  F_{\rm min} (\cW)$ over all encoding unitaries. 

In order to solve Eq.~\ref{eq:fid_loss} each element of $SU(2^n)$ is then parameterized using the so-called \emph{Cartan decomposition}~\cite{khaneja, earp}. The Cartan decomposition provides a way to represent any element of $SU(2^{n})$, upto local unitaries, as a product of operators from $SU(2^{n-1}) \otimes SU(2)$ and unitary operators \emph{nonlocal} on the entire $n$-qubit space. Such a decomposition can be applied recursively, to further decompose each $SU(2^m)$ operator as a product of local and nonlocal unitary operators. 

%and using elements of Abelian subalgebras, %which can be formally stated as follows,
%{\textbf{Cartan decomposition}~\cite{earp}\textbf{.~}} \textit{Any $U \in SU(2^{n})$, for $n > 2$ can be decomposed as,
%\begin{equation}\label{eq:cartan}
%G = K^{(1)} F^{(1)} K^{(2)} J K^{(3)} F^{(2)} K^{(4)}.
%\end{equation}
%where, $K^{(i)}$ are product operators from $SU(2^{n-1}) \otimes SU(2)$, $F^{(j)}\equiv\exp(-\upi f^{(j)})$ and $J\equiv \exp{(-\upi h)}$ are unitary operators \emph{nonlocal} on the entire $n$-qubit space, with $h \in \mathfrak{h}_{n}$ and $f^{(j)} \in \mathfrak{f}_{n}$. Such a decomposition can be applied recursively, to further decompose each $SU(2^m)$ operator in $K^{(i)}$ in the same form as in Eq.~\eqref{eq:cartan}, for all $m=2,3,\ldots,n-1$.}\\

\begin{figure}[H]
\centering
\includegraphics[scale=.45]{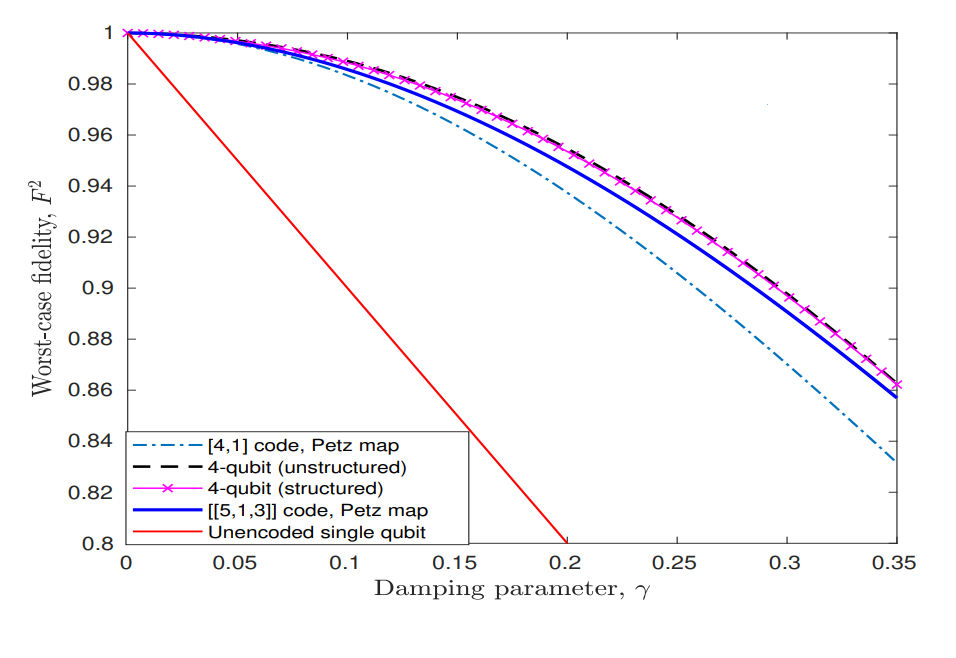}
\caption{ Performance of $4$-qubit noise-adapted codes for amplitude-damping channel, obtained using the numerical search procedure based on the Cartan decomposition.}
\label{fig:4qubit}
\end{figure}

The Cartan form is naturally useful in the context of searching for good codespaces, especially in the regime where the noise assumed to be independent and local. The Cartan decomposition allows one to look for potential codespaces by directly searching over the non-local unitaries in the decomposition, while keeping the local unitaries fixed. For instance, in the case of a $4$-qubit encoding, the total number of parameters required to describe any element in  $SU(2^4)$ is $363$ using the Cartan decomposition.  However, since we only need to search over the intermediate non-local unitaries to find good codespaces, one needs to search over only $82$ parameters. Therefore, this Cartan-form-based search leads to a quick and efficient way of obtaining optimal noise-adapted codes for arbitrary noise channels.

Fig~\ref{fig:4qubit} compares the performance of various quantum codes for the amplitude-damping channel, by plotting the worst-case fidelity as a function of the noise strength $\gamma$. We see that the best $4$-qubit codes obtained via the numerical search outperform the perfect $5$-qubit code (solid line) as well as the noise-adapted $4$-qubit code (dot-dashed line). It further emphasizes the point made earlier that it suffices to search only over the non-local parts of the unitary in the Cartan decomposition (structured search, cross-line), since the best code thus obtained performs comparably to the best code obtained by searching over the entire parameter space in the Cartan decomposition (unstructured search, dashed line).
%Recall that an amplitude damping channel $\cE_\mathrm{AD}$ is described by the following pair of Kraus operators, as, where $\gamma$ is the probability of decay of the excited state $\vert1\rangle$ to the ground state $\vert0\rangle$.

\subsection{Learning-based approaches to construct adaptive QEC codes}

We conclude this section on noise-adapted quantum codes with a survey of some recent works that propose to use learning-based approaches in searching for good QEC protocols.  On the one hand, Ref.~\cite{qvector} demonstrates a variational quantum algorithm to obtain quantum codes specific to the device hardware, in a manner that does not require prior knowledge about the dominant noise model in the system. A simple enough cost function, namely, the average fidelity has been optimized in this algorithm. Alternately, Ref.~\cite{cao2022quantum} develops a variational quantum algorithm to obtain arbitrary quantum codes, not necessarily the channel-adapted ones, but also those with specific code parameters such as distance. In this case, the cost function is defined by generalizing the Knill-Laflamme condition in Eq.~\eqref{eq:perfectqec} and the efficacy of their algorithm is demonstrated by identifying various known codes.

There have also been reinforcement based approaches in finding quantum codes. For example, Ref.~\cite{neural_networks} show how neural network based learning can be used to construct quantum codes for various quantum channels. On a different note, Ref.~\cite{qeccodes} demonstrates a reinforcement-based learning approach to adapting QEC codes by benchmarking them against the desired logical error rate. Finally, Ref.~\cite{reinforcement} demonstrates how to train a network-based agent to learn QEC strategies from scratch and hence  protect a set of qubits from noise.
%\begin{itemize}
%\item \textcolor{blue}{QVECTOR}

%https://arxiv.org/abs/1711.02249
%\item \textcolor{blue}{Reinforcement learning based apparoach}

%https://journals.aps.org/prx/abstract/10.1103/PhysRevX.8.031084
%\item \textcolor{blue}{Quantum codes from neural networks}
%https://iopscience.iop.org/article/10.1088/1367-2630/ab6cdd
%\item \textcolor{blue}{Optimizing Quantum Error Correction Codes with %Reinforcement Learning}
%https://quantum-journal.org/papers/q-2019-12-16-215/

%\item \textcolor{blue}{Mention Markus Grassl's recent work}
%https://arxiv.org/abs/2204.03560
%\end{itemize}

\section{Applications of noise-adapted QEC}\label{sec:applications}

It is well known that ideas from quantum error correction have today permeated diverse fields in physics, from the toric code~\cite{toric_code} which has given rise to several interesting topological models in condensed matter~\cite{baskaran2007} to the more recent connections between stabilizer codes, tensor networks~\cite{ferris2014} and holography~\cite{happy2015}. In this section, we will describe some interesting connections between the theory of approximate or noise-adapted QEC and areas like many body physics and the AdS/CFT correspondence.

\subsection{Quantum state transfer over spin chains}\label{sec:state_transfer}

In the context of quantum many body systems, it was recently shown that approximate quantum codes can occur naturally as ground spaces of certain topologically-ordered, gapped Hamiltonian systems~\cite{elizabeth, elizabeth_2}. Another interesting connection between quantum channels, quantum error correction and many body systems arises in the context of state transfer over spin chains~\cite{bose2007}. Here, we would like to elaborate on this connection and emphasize the role played by approximate QEC in achieving information transfer over spin chains with a high degree if fidelity.

Quantum communications using interacting spins is one of the areas that has generated a lot of interest over the recent years. Following the pioneering work of Bose~\cite{bose} who showed that such a state transfer could be viewed as transmitting quantum information over a quantum channel, multiple protocols demonstrating \emph{perfect} and \emph{pretty-good} state transfers using spin chains have been proposed. Rather than use a single spin chain, it was subsequently shown that conclusive and perfect state transfer could be achieved using a pair of spin-$1/2$ chains with dual-rail encoding~\cite{Burgarth}. As opposed to a perfect transfer, in the case of pretty good transfer one finds an optimal scheme to transfer the information with high fidelity, using permanently coupled spin chains~\cite{osborne}. Going beyond the $2$-qubit dual-rail code, it has been shown that state transfer can be achieved over disordered spin chains beyond the localization length, by using a concatenated form of the standard $5$-qubit code~\cite{allcock}. Alternately, the role of QEC in enhancing fidelity of state transfer under local noise affecting each individual spin in the chain has also been studied~\cite{kay_2016}.

Since the underlying quantum channel that arises in the context of state transfer over spin chains is typically not of the Pauli form, a natural question to ask is whether  noise-adapted QEC may enable information transfer using spin chains over longer distances. This question was addressed recently~\cite{ak_statetransfer}, where it was shown that pretty good state transfer can be achieved using adaptive QEC protocols on spin chains. We will now briefly describe such a noise-adapted protocol for quantum state transfer over a $1$-d Heisenberg chain. 

Consider the problem of transmitting a single qubit worth information across a spin-$1/2$ chain with the interaction Hamiltonian given by,
\begin{equation} \label{eq:H_gen}
\cH = -\sum_{k} J_{k}\left(\sigma^{k}_{x}\sigma^{k+1}_{x}+\sigma^{k}_{y}\sigma^{k+1}_{y}\right) - \sum_{k}\tilde{J}_{k}\sigma_{z}^{k}\sigma^{k+1}_{z} + \sum_{k}B_{k}\sigma_{k}^{z}, 
 \end{equation}
 where,  $\{J_{k}\}>0$ and $\{\tilde{J}_{k}\}>0$ are site-dependent exchange couplings of a ferromagnetic spin chain, $\{B_{k}\}$ denote the magnetic field strengths at each site, and, $(\sigma^{k}_{x},\sigma^{k}_{y},\sigma^{k}_{z})$ are the Pauli operators at the $k^{\rm th}$ site. 
The standard state transfer protocol works by initialising a single spin at site $s$ (\emph{sender}) to the quantum state to be transferred and then \emph{receiving} it at site $r$. It is easily shown that if the spins are interacting via the Hamiltonian in Eq.~\eqref{eq:H_gen}, this leads to a quantum channel $\cE$ from the input spin to the  receiver's end with Kraus operators,
\begin{equation}
E_{0}  = \left( \begin{array}{cc}
1 & 0 \\
0 & f_{r,s}^{N}(t)
\end{array} \right), \; E_{1} = \left( \begin{array}{cc}
0 & \sqrt{1-\vert f_{r,s}^{N}(t)\vert^{2}} \\
0 & 0
\end{array} \right). \label{eq:Kraus_ideal}
\end{equation}
The Kraus operators in Eq.\eqref{eq:Kraus_ideal} lead to a channel that has the same structure as the amplitude-damping channel described in Eq.~\eqref{eq:ampdamp}, but is more general since the parameter $f_{r,s}^{N}(t)$ characterizing the noise in the channel is complex. The parameter $f_{r,s}^{N}(t)$ is the so-called \emph{transition amplitude}, $r$ refers to the receiver's site and $s$ refers to the senders site and $N$ refers to the total number of spins on the spin chain. %and it satisfies,
%\begin{eqnarray} \sum_{r=1}^{N}\vert f_{r,s}^{N}(t)\vert^{2} &=& 1, \, \forall \; s = 1,2,\ldots, N . \nonumber \\
%\sum_{k=1}^{N}f^{N}_{r,k}(t)(f^{N}_{k,s}(t))^{*} &=& \delta_{rs} , \, \forall \; k = 1,2,\ldots, N . \label{eq:trans_amp}
%\end{eqnarray}
%where $\delta_{rs}$ is the delta function with $\delta_{rs} = 1 $ for $r = s$ and $\delta_{rs} = 0$ for $r\neq s$.

\begin{figure}[H]
\centering
\includegraphics[scale=.3]{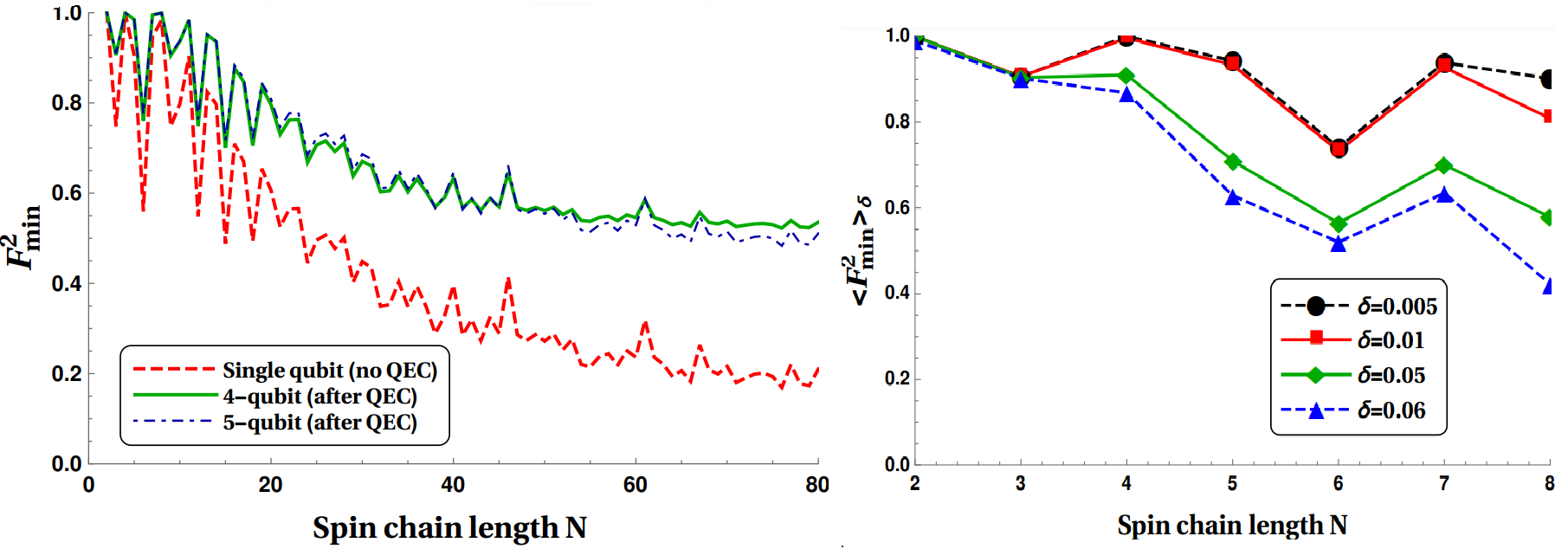}
\caption{a) Spin transfer on an ideal XXX chain (left). b) Spin transfer on a disordered XXX chain (right) }
\label{fig:spinchain}
\end{figure}

Since the quantum channel $\cE$ looks very similar to amplitude-damping channel, the adaptive QEC protocol in~\cite{ak_statetransfer} picks the $4$-qubit code in Eq.~\eqref{eq:4qubit} and the Petz recovery map in Eq.~\eqref{eq:Petzmap} adapted to this code to achieve state transfer with pretty good fidelity. The plots in Fig.~\ref{fig:spinchain} describes the performance of such an adaptive QEC protocol on a spin chain of length $N$ subject to the $XXX$ interaction, with $\{B_k=0\}$ and $\{J_i=J\}$ in Eq.~\eqref{eq:H_gen}.

In these plots, the performance of the state transfer protocol from the first to the $N^{\rm th}$ site is characterised using the worst-case fidelity as function of the length $N$ of the spin chain. The first plot in Fig.~\ref{fig:spinchain} shows the state transfer fidelities for an ideal $XXX$ chain, whereas the second plot shows the fidelities obtained for disordered spin chain with disorder strength $\delta$. In this case, the disorder-averaged worst-case fidelity $\langle F_{\rm min}^2\rangle_\delta$ is the performance metric, and the noise-adapted Petz recovery is constructed in terms of the disorder-averaged transition amplitude $\langle f_{r,s}^{N}(t)\rangle_\delta$. It is shown both analytically as well as numerically that for small disorder strengths the disorder-averaged transition amplitude is the same as the no-disorder case~\cite{ak_statetransfer} and the variance in the transition amplitude is vanishingly small. This makes a convincing case for using the noise-adapted Petz recovery to achieve pretty good state transfer over ideal as well as disordered spin chains. 

%\subsection{Approximate QEC in many body systems}

\subsection{The Petz map in Ads/CFT}

Moving on from many body systems to holography, we now provide a brief summary of the interesting emerging connections between perfect/approximate quantum codes and the bulk-boundary correspondence in Ads/CFT. The AdS/CFT duality proposes a correspondence between a theory of quantum gravity on a $d+1$-dimensional (asymptotically) anti-deSitter (AdS) spacetime and a conformal field theory (CFT) in one less spatial dimension defined on its boundary. It was shown in the early works of Almheiri \emph{et al}~\cite{almheiri2015} and Pastawski \emph{et al}~\cite{happy2015, pastawski2017} that the framework of quantum error correction provides an elegant characterization of this bulk-boundary correspondence. The mathematical formalism of QEC is naturally suited to answer the question of bulk reconstruction, namely, can one find a representation of operators in the bulk gravity theory as operators acting on a subregion of the boundary. 

Viewing the AdS/CFT correspondence as a map from the bulk to the boundary, a noisy quantum channel arises from tracing over a subregion of the boundary. The problem of bulk reconstruction while having access only to a subregion of the boundary, is then the same as the problem of recovering from this erasure noise channel. Formally, it has been shown that such a reconstruction is possible when there is an \emph{exact} equivalence of the bulk relative entropy and boundary relative entropy~\cite{dong2016, jlms2016}. In practice one can only expect \emph{approximate} equality of the bulk and boundary relative entropies, and this naturally leads to the question of whether approximate or noise-adapted recovery maps like the Petz map might find use in the context of the bulk-boundary reconstruction. 

We will conclude with a quick survey of some recent results in this area and refer to~\cite{holography_qec} for a detailed review of this emerging field. The first concrete proposal for an approximate recovery map that could solve the problem of bulk reconstruction involved a variant of the Petz map, often called the \emph{twirled} Petz map~\cite{cotler2019}. Recall that the original state-dependent Petz map was originally derived as the recovery map that characterizes saturation of monotonicity of relative entropy under a noise channel. The twirled Petz map is a generalization of the original Petz construction which was shown to characterize approximate saturation of the monotonicity of relative entropy under a noise channel~\cite{junge2018}. Subsequently, it was shown that approximate reconstruction of the bulk operators can be achieved via an averaged version of the Petz map~\cite{chen2020}. In related work, a Petz-based reconstruction map was demonstrated for toy models of holography based on random tensor networks~\cite{jia2020petz}. Ultimately, a complete understanding of the role of approximate recovery maps in the holographic setting requires a construction of such maps for general von Neumann algebras. There has been some progress in this regard recently, with a modified Petz map for arbitrary von Neumann algebras~\cite{faulkner2022}, but the quest is on to find a suitable physical construction of a universal, approximate recovery map that can solve the bulk reconstruction puzzle in holography. 

\section{Noise-adapted techniques for quantum fault tolerance}\label{sec: FT}

A practical quantum computer suffers from decoherence due to the inevitable system-environment interactions. The theory of fault-tolerant quantum computing~\cite{preskill1998} provides a framework by which reliable and long computations can be done even in the presence of noise, which could arise either from gate imperfections or noisy quantum systems. A fault-tolerant computation proceeds by first encoding the information into a quantum error correcting code and then performing encoded gate operations on them in a noise-resilient way. The errors generated are then to be removed periodically, before they can cause irreparable damage. 

Central to this theory is the \emph{threshold theorem} which provides a critical threshold value, namely, the noise strength below which fault tolerance scheme is successful and the encoded operations outperform the unencoded operations~\cite{aliferis2006}. This threshold number depends crucially on the QEC protocol being used and the noise model assumption. For example, Ref.~\cite{cross2009} provides a comparative study of the threshold values achieved by different quantum codes for the case of depolarizing noise, wheres Ref.~\cite{campbell2017} discusses the thresholds achievable by topological codes such as the surface codes. The threshold numbers obtained thus far against various noise models by specific quantum codes are tabulated in Table.~\ref{tab:table1}. 

As with standard QEC, much of the work on fault tolerance (FT) in the past has been based on \emph{perfect} quantum codes that are essentially based on Pauli noise models.  However, in most cases, the dominant system noise is non-Pauli, and in these cases the standard fault-tolerance scheme is too resourceful to be implemented on current NISQ hardware. Recently, there have been efforts towards developing noise-adapted fault tolerance schemes that are tailored to the dominant noise processes in physical systems, and these will form our focus in the following sections.

\begin{table}
%\hspace*{-1cm}
\begin{tabular}{|c|c|c|}
\hline
Fault-tolerant Scheme& Noise model& Threshold\\
\hline
\hline
General&Simple stochastic noise&$\sim 10^{-6}$\\
\hline
Polynomial codes (CSS codes)&general noise model&$10^{-6}$\\
\hline
Concatenated $7$-qubit code&Adversarial stochastic noise&$2.7 \times 10^{-5}$\\
\hline
Error detecting concatenated codes ($C_4/ C_6$)&depolarizing noise&$1\%$\\
\hline
Concatenated $7$-qubit code&Hamiltonian noise&$10^{-8}$\\
%\cline{2-4}
\hline
Concatenated repetition code& biased dephasing noise & $0.5 \%$, $0.24\%$ \\
\hline
Surface codes&depolarizing&$0.75\%$ , $0.5\% -  1\%$\\
\hline
$[[4,2,2]]$ concatenated toric code& depolarizing&$0.41\%$\\
\hline
\end{tabular}
\vspace{.4cm}
\caption{\label{tab:table1} Noise thresholds for various quantum fault tolerance schemes~\cite{thesis}.}
\end{table}

%\begin{itemize} \item \textcolor{blue}{Review past work on FT against biased-noise models}\item \textcolor{blue}{Put in a table comparing different FT thresholds for different codes and noise models?}

%\end{itemize}

\subsection{Quasi-exact fault tolerance}
One interesting approach that goes beyond the standard prescription of quantum fault tolerance (FT) is the theory of \emph{quasi-exact fault tolerance}~\cite{wang,quasi}. This scheme demonstrates fault tolerance using a class of approximate quantum codes called \emph{quasi codes}. These are approximate quantum codes  defined in terms of scaling parameters, which when tuned can lead to exact codes in some limit. The advantage of this framework of quasi-exact does is that it allows one to interpolate between approximate and perfect codes and correspondingly define parameters like (quasi) code distance which cannot be defined for other classes of approximate codes. Examples of quasi codes include valence-bond solid codes and symmetry-protected topological states that occur naturally in quantum many body systems. 

Quasi-exact FT further invokes a weaker version of universality referred to as \emph{quasi universality}. This is a weakening of the usual universality, which induces a coarse-graining structure on the unitary group by identifying a suitable cut-off. Specifically, a set of logical unitary operators, $U$ $\in$ $SU(d)$ is now realized as a coarse-grained unitary $U'$ $\in$ $SU(d)_\eta$, where $\eta$ is the accuracy, which is the distance between $U'$ and $U$, and $SU(d)_\eta$ $\in$ $SU(d)$. One interesting aspect of quasi-exact FT is that it opens up the possibility of circumventing the no-go theorem for regular FT, namely, that transversality and universality cannot coexist for exact codes~\cite{eastin}. The theory of quasi-exact codes and quasi-exact fault tolerance does allow for transversality with \emph{quasi-universality}, but, the computations allowed by quasi-exact FT are only finite length. The notion of quasi-exact FT is strictly weaker than the standard fault tolerance since the uncorrectable errors accumulate with the number of operations performed.

%\begin{itemize}

%\item\textcolor{blue}{https://iopscience.iop.org/article/10.1088/1367-2630/ac4737/meta}
%\item\textcolor{blue}{https://journals.aps.org/prresearch/abstract/10.1103/PhysRevResearch.2.033116}

%\end{itemize}

\subsection{Achieving fault tolerance using noise-adapted quantum codes}

Much of the work on quantum fault tolerance assumes that the noise is of the Pauli type and aims to build encoded units that are tolerant against Pauli-type faults. Even in cases where the fault tolerance scheme is tailored to a specific noise model, the dominant noise has been typically of the Pauli type. For instance, Ref.~\cite{biased_noise} presents a fault-tolerant scheme specific to dephasing noise using the $3$-qubit phase flip code. In contrast to such approaches, a recent work demonstrates the possibility of biased-noise fault-tolerance based on a noise model as well as QEC code that is non-Pauli~\cite{ak_ft}. Specifically, this work assumes that the dominant noise affecting the physical qubits is amplitude-damping channel. Expanding the amplitude damping channel in Eq.~\eqref{eq:ampdamp} in the limit of small values of the noise
parameter $p$,
\begin{equation}\label{eq:ampdamp2}
\cE_\mathrm{AD}(\,\cdot\,) =\tfrac{1}{4}{\left(1+\sqrt{1-p}\right)}^2\cI(\,\cdot\,)+p\cF(\,\cdot\,)  +{\left[\tfrac{1}{4}p^2+O(p^3)\right]}Z(\,\cdot\,)Z,
\end{equation}
where $\cI(\cdot)\equiv (\cdot)$ is the identity channel, and $\cF(\cdot)$ is the TP (but not CP) channel defined by,
\begin{equation}\label{eq:amp_fault}
\cF(\,\cdot\,) \equiv\tfrac{1}{4}{\left[(\,\cdot\,)Z+Z(\,\cdot\,)\right]}+E(\,\cdot\,)E^{\dagger} \equiv \tfrac{1}{2}\cF_z(\,\cdot\,) + \cF_a(\,\cdot\,).
\end{equation}
In Eq.~\eqref{eq:amp_fault}, $\cF_Z$ refers to an off-diagonal error leading to a non-state operator and $\cF_a$ refers to a damping error.

Following the basic principles of quantum fault tolerance developed in~\cite{aliferis2006}, Ref.~\cite{ak_ft} constructs encoded units tolerant against faults of upto $O(p)$ in Eq.~\eqref{eq:ampdamp2} arising from amplitude-damping noise. Starting with faulty physical units, and using the $4$-qubit code given in Eq.~\eqref{eq:4qubit}, it describes a fault-tolerant universal encoded gate set comprising of $\{S, T, H, \textsc{cphase}, \textsc{ccz}\}$ and a fault-tolerant error correction unit. Furthermore, going against the conventional wisdom of quantum fault tolerance, this work shows that there could be transversal gates that are not fault tolerant, such as the \textsc{cnot} gate in the case of amplitude-damping noise. The transversal \textsc{cphase} gate turns out to be fault-tolerant to single faults arising from ampltiude-damping noise. Thus, \textsc{cphase} gate is a \emph{noise-structure preserving} gate, which is then used to implement the set of other single-qubit encoded gates via teleportation. We note here that a similar idea of identifying noise-bias-preserving gates has been recently discussed in the context of systems that are subject to biased Pauli noise~\cite{puri2020bias, xu2022}. 

\begin{figure}[H]
\centering
\includegraphics[scale=.6]{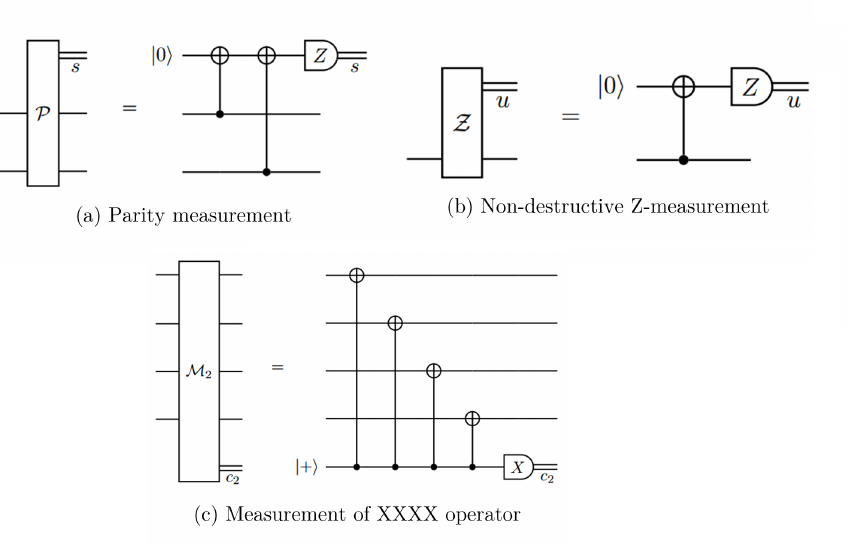}
\caption{Basic units of error correction using approximate $4$-qubit code.}
\label{fig:ec_unit}
\end{figure}

Finally, we briefly discuss the nosie-adapted recovery circuit used to construct the fault-tolerant error correction (\textsc{ec}) unit in~\cite{ak_ft}. Since the $4$-qubit code is approximate and adapted to a non-Pauli kind of noise, in order to perform QEC one has to measure more than just the stabilizer set defining the codespace. Specifically one needs to measure the set of two qubit nearest neighbour $Z$ operators— $\{ZZII, IIZZ\}$ and single qubit $Z$ operator — $\{ZIII\}$ in order to find exactly where the damping has occurred. Finally one measures $XXXX$ to restore the superposition and to kill the off-diagonal errors. The basic units required to build the error correction unit are shown in Fig.~\ref{fig:ec_unit}. These units are then strung up fault-tolerantly~\cite{ak_ft} with ancilla qubits and flag qubits to build a fault-tolerant \textsc{ec}-unit. These fault-tolerant circuit constructions lead to an estimation of a pseudo-threshold value for amplitude-damping noise. This was estimated as $5.13\times 10^{-5}$ for the memory unit and $2.26\times 10^{-5}$ for the encoded \textsc{cz}-unit~\cite{ak_ft}.

 %It has been noted that both the two-qubit encoded gates, namely, \textsc{cnot} as well as \textsc{cz} are transversal, however, only \textsc{cz} encoded gate is fault tolerant. Contrasting to a \textsc{cnot} physical gate, a \textsc{cz} gate propagates errors which the $[[4,1]]$ code can correct. Therefore, the transversal \textsc{cz} gate is a \emph{noise-structure preserving gate}, with which other encoded, single-qubit gates in the universal gate set are realized via teleportation. 

%noise-structure-preserving gates \subsection{Other aspects of Approximate/Adaptive QEC}

\section{Summary and Outlook}\label{sec:summary}

The current era of NISQ devices presents both a challenge and an opportunity for theorists and experimentalists alike, from the perspective of error correction and fault tolerance. On the one hand, in order to fully tap into the potential of the available quantum technologies, one has to ascend the ladder of \emph{scalability} and \emph{control} over the qubits simultaneously~\cite{jurcevic2021}. On the other hand, these NISQ devices also provide an opportunity to rework our theoretical frameworks and perform experiments that leverage NISQ devices to test theoretical ideas. These include, for instance, experiments on the quantum cloud that validate the theory of quantum error correction~\cite{pokharel2018demonstration, ghosh2018}, while also providing some insight into the nature of the dominant noise process in the system~\cite{repibm} and %the hardware ~\cite{zulehner,wangibm}, 
studies that demonstrate applications to quantum chemistry (see~\cite{nisq_review} for a recent review) and nuclear structure~\cite{dumitrescu}. %One experiment validating the existing theory of QEC by implementing a $15$- qubit repetition code. Ref. also studies the dominant noise affecting the system. The progress with computations on NISQ devices could speedup if one performed tasks specific to the system hardware.

In this review, we have surveyed QEC approaches that deviate from the standard QEC and fault tolerance formalism, and instead aim to correct for the dominant noise affecting the quantum devices under consideration. Such noise-adaptive QEC strategies are known to be less resourceful~\cite{leung,fletcher_AD,hui_prabha}, but they also require non-trivial circuit constructions~\cite{gilyen2022} and syndrome extractions~\cite{ak_ft} in contrast to the simpler Pauli measurements for standard stabilizer codes. Thus, an immediate question to address is whether such noise-specific techniques can be scaled up to be able to suppress more errors and perform long computations reliably. One possible approach towards such a scale up could be to concatenate a noise-adapted code tailored to the dominant noise in the system with another QEC code of the same or different type. Alternately, one could use channel-adapted techniques on lattice-based codes like the Bacon-Shor code~\cite{piedrafita2017reliable} and surface codes which have the ability to tolerate more errors depending on their lattice size.

The NISQ era has also spawned the development of newer ideas such as \emph{quantum error mitigation}~\cite{cao2021nisq} which focuses on the limited goal of reducing the effective noise levels in near-term quantum devices, rather than implementing a complete QEC protocol. This in turn opens up the possibility of merging such hardware-specific error mitigation strategies from standard QEC and fault tolerance~\cite{qem2022}. Going forward, efficient and powerful optimization techniques could lead us to better adaptive QEC strategies with optimal encodings and recovery. For instance, optimization involving machine learning~\cite{reinforcement} and other learning-based approaches including a quantum variational strategy~\cite{cao2022quantum} are already being explored. 

In the coming years, we expect to realize qubits with longer coherence times and gates with higher accuracy in the lab. In the meantime, learnings from noise-adapted QEC and fault tolerance will continue to give us insights on how well we can make use of NISQ devices, and develop strategies that will enable us to make the transition from noisy quantum devices to robust and scalable quantum computing architectures.

\section{Acknowledgements}
This work is supported in part by a Seed Grant from the Indian Institute of Technology Madras, as part of the Centre for Quantum Information, Communication and Computing. P.M. acknowledges financial support by the Department of Science and Technology, Govt. of India, under Grant No. DST/ICPS/QuST/Theme-3/2019/Q59. 
\section{Declarations}
Funding and/or Conflicts of interests/Competing interests: On behalf of both authors, the corresponding author states that there is no conflict of interest. 
\bibliography{review}

\end{document}